\documentclass[11pt]{article}
\usepackage{moriond2000,epsfig}

\def\Journal#1#2#3#4{{#1} {\bf #2}, #3 (#4)}

\def\be{\begin{equation}}
\def\ee{\end{equation}}
\def\bea{\begin{eqnarray}}
\def\eea{\end{eqnarray}}

\begin{document}
\vspace*{4cm}
\title{A GRAVITATIONAL LENS SURVEY WITH THE {\it PLANCK SURVEYOR}} 

\author{Andrew W. Blain }

\address{Institute of Astronomy, Madingley Road, Cambridge, CB3 0HA, UK} 

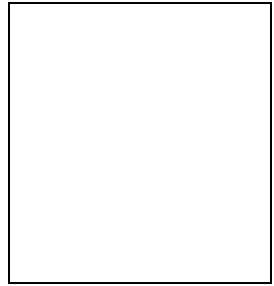
\maketitle\abstracts{
The {\it Planck Surveyor} cosmic microwave background (CMB) imaging mission 
will make very sensitive maps of the whole sky at microwave, millimetre
(mm) and sub-mm wavelengths. The steep source counts expected in the 
highest frequency {\it Planck} bands are likely to be associated with a strong 
magnification bias, and so the fraction of galaxies magnified by a factor of two or 
more could be greater than 10\%. In this paper, previous predictions of the 
significance of the {\it Planck} survey for studies of lensing statistics are updated, to 
reflect our expanded knowledge of the properties of high-redshift dusty 
galaxies, obtained from far-infrared and sub-mm surveys. A catalogue of 
probably about $10^{4}$ galaxies, but perhaps as many as 10$^5$ or as few as 
$10^3$, is expected to be generated from the final {\it Planck} all-sky maps. Of order
1000 galaxies, a better determined number, might reasonably be expected 
to be strongly lensed. Co-ordinated sub-mm and far-infrared follow-up observations 
made using the SPIRE and PACS instruments aboard the ESA cornerstone mission 
{\it FIRST} -- {\it Planck}'s travelling companion to its L2 orbit -- and the ground-based 
ALMA interferometer array will provide accurate positions and very valuable 
information about the spectral energy distributions (SEDs), redshifts and
astrophysical properties of the galaxies 
in the {\it Planck} catalogue. ALMA and radio images will also reveal their 
morphology, and the characteristic arc and multiple image structures produced 
by gravitational lensing. Studies of these bright sub-mm-selected galaxies will 
allow new insight into the process of galaxy formation and evolution. 
}

\section{Introduction}

The {\it Planck Surveyor} CMB imaging satellite\cite{Bersanelli} will provide a 
mm/sub-mm-wave map of the whole sky with a resolution of about 5\,arcmin down to 
a detection limit of order 100\,mJy at wavelengths of 
350, 550 and 850\,$\mu$m. These wavelengths are expected to be close to the peak of 
the redshifted thermal dust emission spectrum of distant galaxies. The effect of 
redshifting the dust spectrum peak into these observing bands is to 
provide access to the distant Universe, with little contamination 
from bright low-redshift galaxies. This remarkable, helpful $K$-correction is 
confirmed observationally to be effective in the sub-mm waveband, and is an almost 
unique feature of the waveband, although a similar effect may be at work in the hard 
X-ray surveys now being carried out. A priori, the counts of sub-mm galaxies at the 
faintest depths probed by {\it Planck} are 
expected to be very steep, an expectation which seems to be confirmed by the results 
of recent ground-based sub-mm-wave surveys.\cite{B_ALMA} 

Very steep sub-mm counts generate a strong positive magnification bias,\cite{B96,B98}  
and thus boost the fraction of lensed galaxies in a flux-limited sample. This 
enhancement to the surface density of lensed galaxies will hopefully be exploited 
in the {\it Planck} survey to allow a very large catalogue of mainly unknown strongly 
lensed galaxies to be compiled.  

\section{Sensitivity and confusion noise in a {\it Planck} survey}

The expected sensitivities of the {\it Planck}-HFI instrument in its three highest 
frequency channels, in which the largest number of distant dusty galaxies and lenses 
will be detected,\cite{B98} are listed in Table\,1, along with the predicted levels of 
source confusion noise.

It is important to understand the significance of source confusion noise 
introduced into the {\it Planck} images due to the varying number and 
flux densities of unresolved dusty galaxies in the observing beam. 
At 5\,arcmin across, the {\it Planck} beam is very much larger than the mean
separation of $L^*$ galaxies, and also about 4 times greater in area than the 
5-arcmin$^2$ field of view of the SCUBA camera at the JCMT, from which most 
observational information about the sub-mm source population has thus far been 
derived. 

In Fig.\,1 the results of many independent simulations of sampling a random 
unclustered distribution of sources on the sky with the observing beam in the 
relevant {\it Planck}-HFI channels are compared with the estimates of 
instrumental noise. Confusion noise due to Galactic cirrus predicted by 
{\it IRAS}-based studies,\cite{Boulanger} and of simple estimates based on 
the surface density of galaxies,\cite{BIS,B_Carnegie} are compared in Table\,1. 
All these calculations of the confusion signal due to extragalactic sources are 
made assuming a model for the underlying surface density of sources on the sky that 
is at the top end of the range of possible values, the predictions are therefore likely 
to be conservatively high, and thus pessimistic for making reliable detections. 
Extragalactic confusion noise is expected to dominate both the instrumental and 
Galactic noise in these bands. 

In Fig.\,1, the results of the simulations are enveloped by solid curves, which 
represent log-normal distributions providing good representations of the 
results. The value of $\sigma$ in the log-normal distribution 
($\exp - [{\rm ln}x - \bar x]^2/2 \sigma^2$) at all three frequencies is close to 0.2. 

Unfortunately, the actual level of confusion noise that will be contributed to the 
{\it Planck} image depends on the uncertain surface density of 
galaxies that are brighter than those selected in existing sub-mm surveys (several 
tens of mJy at 850\,$\mu$m/350\,GHz). The results of future wider-field ground-based 
sub-mm surveys, and experience gained from the results of long-duration 
balloon-borne mm- and sub-mm-wave CMB experiments, including the existing 
BOOMERanG data and forthcoming data from TOPHAT will be useful in this respect.
Recent simulations of extraction algorithms,\cite{Hobson} in which the surface density of
galaxies is assumed to be a factor of 2-3 times lower than that assumed 
here,\cite{Toffolatti} suggest that 350-GHz point sources with flux densities greater 
than of order 75\,mJy could be extracted from the {\it Planck} all-sky map.

\section{Expected source density and numbers of detections} 

The two very different counts of unlensed galaxies shown in Fig.\,2 should provide an  
envelope to the maximum range of possible values of the sub-mm counts at flux 
densities at which galaxies will be detected using {\it Planck}. 
The two underlying models\cite{BSIK,BJSLKI} predict very 
different bright sub-mm counts, but are both consistent with the observed far-infrared 
and sub-mm-wave counts and background radiation intensity, and with what little is 
currently known about the redshift distribution of sub-mm-selected galaxies.
Our knowledge of the population will continue to improve 
until the launch of {\it Planck}. Some of the facilities that will contribute to 
this knowledge are listed elsewhere.\cite{B_Carnegie} 

Current observational limits to the population of sub-mm galaxies are reasonably 
well determined at flux densities less than about 
10\,mJy at 350\,GHz from the results of several 
independent surveys: the surface density of galaxies brighter than 4\,mJy at 
850\,$\mu$m/350\,GHz is about 2000\,deg$^{-2}$. These surveys have mainly been 
carried out using the SCUBA camera at the JCMT, but results are also now coming in 
from the MAMBO 1.25-mm bolometer array camera at the IRAM 30-m 
telescope.\cite{Carilli2000} These are flux densities considerably fainter than {\it Planck} 
will probe, and so far,  these instruments have mapped only very small regions of sky 
(several hundred square arcminutes). The counts of objects brighter than 
about 10\,mJy has been only weakly constrained. 

Information is also available about the 
counts at the very brightest flux densities, based on the results of a targeted SCUBA 
survey of low-redshift galaxies selected from the {\it IRAS} catalogue.\cite{Dunne} 
This survey was used to define a luminosity function of local {\it IRAS} galaxies
in the sub-mm, which can be used to impose a lower limit to the bright counts at flux 
densities brighter than the detection limit in the all-sky {\it Planck} survey: this lower 
limit is about 10 sources on the sky brighter than 1\,Jy at 850\,$\mu$m/350\,GHz.

The probability of lensing by foreground galaxies out to the high redshifts, 
at which a significant fraction of the sources detected in the {\it Planck} 
survey are expected to lie, is reasonably well-known, certainly to within a factor of a few. 
Specific 
predictions of this probability, tailored to observations in the sub-mm waveband, are 
discussed elsewhere.\cite{B98,BMM} The predicted density of lenses on the sky should 
also be quite well determined at the flux densities that will be probed by {\it Planck}. 
This is because the intrinsic flux densities of these objects, after correcting for the 
effect of lensing, are likely to be of order 10\,mJy. At this flux density the surface 
density of galaxies is reasonably well constrained by SCUBA 
observations.\cite{B_Carnegie} Predicted numbers of lensed objects\cite{BMM} 
are shown by the dot-dashed lines in Fig.\,2.  

\begin{table}[t]
\caption{
Some approximate parameters describing the likely properties of a {\it Planck} 
all-sky survey that are relevant to the detection and study of dusty galaxies. 
The estimated final sensitivity of the {\it Planck} survey$^1$
$\sigma_{\rm sens}$, and three estimates of confusion values are listed:
that expected from Galactic cirrus$^{11}$ in regions with a 100-$\mu$m 
surface brightness $B_0=1$\,MJy\,sr$^{-1}$, $\sigma_{\rm cirrus}$, a simple 
estimate of the value expected due to external galaxies -- the 
flux density at which the surface density of galaxies exceeds 
1\,beam$^{-1}$,$^{6,4}$ -- $\sigma_{\rm beam}$, and the width of the 
distributions shown in Fig.\,1, between which 65\% of the results of the 
simulations lie, $\sigma_{\rm sim}$. 
Note that the confusion noise predicted by the simulations is log-normal and 
not Gaussian. 
}
\vspace{0.4cm}
\begin{center}
\begin{tabular}{|c|c|c|c|c|}
\hline
$\nu$/GHz & $\sigma_{\rm sens}$/mJy & $\sigma_{\rm beam}$/mJy & 
$\sigma_{\rm cirrus}$/mJy & $\sigma_{\rm sim}$/mJy \\
\hline
353 & 16 & 10 &  1.3 & 40 \\
545 & 19 & 30 & 9 & 60 \\
857 & 26 & 63 & 55 & 170 \\
\hline
\end{tabular}
\end{center}
\end{table}

A significant potential uncertainty remains in these results, however, because 
the maximum magnification that can be produced by a lens is limitted by the size
of the source, being smaller for larger sources. In these calculations, we have assumed 
that background high-redshift sources are less than about 10\,kpc in size, so 
that a maximum magnification of several tens can be produced. This assumption is 
supported by high-resolution interferometric observations of the size 
of the continuum emitting region in low-redshift ultraluminous galaxies, 
which is found to be much smaller -- several 100\,pc across.\cite{Solomon} However, 
there are indications that at least some very luminous dusty 
high-redshift galaxies display CO and dust emission on scales 
greater than 10\,kpc.\cite{Papa,NewFrayer} Unfortunately, sub-arcsecond 
resolution images are required in order to measure the sizes of high-redshift 
dusty galaxies smaller than 10\,kpc, and the necessary interferometric 
observations are difficult and time-consuming. Hence, as we do not yet 
know the size distribution of these distant ultraluminous galaxies, we cannot 
be certain that their sizes are typically small enough for large lensing 
magnifications to be possible. Despite this caveat, individual examples of  
high-redshift sub-mm objects lensed by foreground galaxies with magnifications 
of several tens are known,\cite{10214,APM,Cloverleaf} and so it is 
likely that {\it Planck} will be able to detect a significant number. 

Based on an extrapolation of the log-normal distribution of pixel values predicted by
the conservative simulations of confusion noise shown in Fig.\,1, it is possible to 
predict the number of pixels that are likely to exceed a certain flux density due to 
the effects of confusion over the all-sky {\it Planck} survey. This expected count of 
spurious detections, which is shown by the dotted lines in Fig.\,2, can then be 
compared with the expected count of lensed and unlensed galaxies. Reliable detections  
should be possible if these counts lie safely above the dotted lines at any flux density. 
Hence, it is likely that lensed galaxies brighter than about 250, 500 and 1500\,mJy 
at 353, 545 and 857\,GHz respectively could be detected using {\it Planck}, 
even in this pessimistic model of the confusion noise. At these limits, of order 
100 lensed sources are expected over the whole sky. If the counts of bright unlensed 
sources are less than assumed here,\cite{Hobson} then the confusion noise will 
be reduced, and a greater number will be 
detected. The counts shown in Fig.\,2 allow these different scenarios to be 
investigated directly. It is important to note that much better estimates of 
confusion noise will be available well in advance of the launch of {\it Planck}. 

The emission from both lensed and unlensed distant galaxies 
will pass unattenuated through the Galactic plane in the {\it Planck} bands; however, 
the ability to distinguish them against a bright and structured Galactic foreground is 
likely to be limited. The level of Galactic confusion (see Table\,1) depends on the 
Galactic background intensity $B_0$ as $B_0^{1.5}$.\cite{Boulanger} The
results suggest that the sensitivity of the {\it Planck} survey can only be exploited 
fully to find extragalactic sources in regions of the sky where the 100-$\mu$m 
surface brightness from the Milky Way is less than about 5\,MJy\,sr$^{-1}$, 
a condition which is satisfied over most of the sky.\cite{SFD} 

\begin{figure}
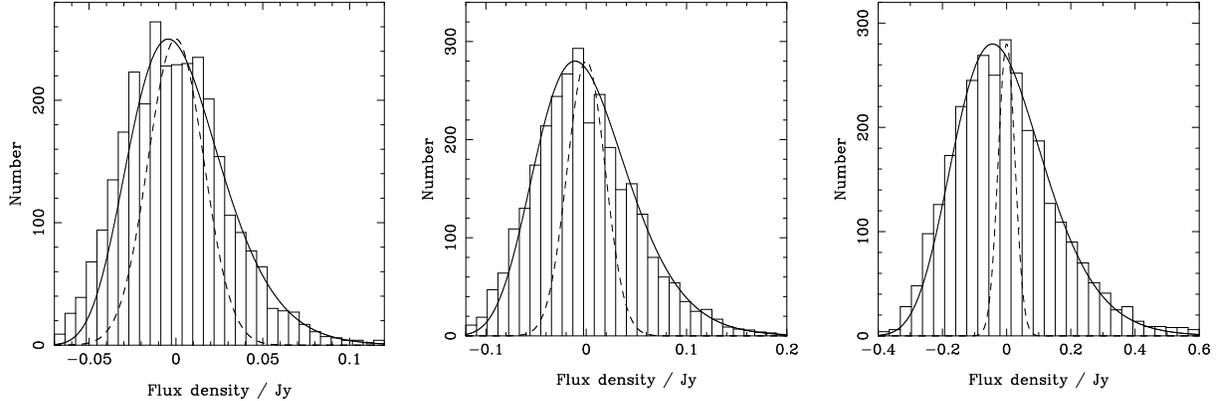

\psfig{figure=Conf_353_rev.ps,height=1.97in,angle=-90}
\hskip 10pt
\psfig{figure=Conf_545_rev.ps,height=1.97in,angle=-90}
\hskip 10pt 
\psfig{figure=Conf_857_rev.ps,height=1.97in,angle=-90}
\caption{Results of simulations of confusion noise expected in the 
5-arcmin beam of the {\it Planck}-HFI instrument at 
frequencies of 353\,GHz (left), 545\,GHz (middle) and 857\,GHz (right). 
The histograms show the results of 3000 different simulation realisations of 
the predicted galaxy distribution$^8$ sampled in 
in the {\it Planck} beam, with Gaussian instrumental noise added. 
The solid curves show analytical log-normal distributions that adequately 
describe the envelopes of the simulation results. The dashed curves show the 
expected Gaussian instrumental noise in the {\it Planck} 14-month 
survey (16, 19 and 26\,mJy\,beam$^{-1}$ respectively). At  
frequencies lower than 353\,GHz instrumental noise is 
expected to dominate confusion 
noise, a trend which can be seen developing from frequency to frequency above. 
The flux scale has been shifted to give a zero net background. At 353\,GHz 
Galactic foreground 
confusion noise is not included, as this source of noise is not expected to be 
more significant than the instrumental noise in regions of low Galactic emission.} 
\end{figure}

\section{Follow-up observations} 

From the {\it Planck} all-sky images alone, point sources will be detectable, but 
no information will be available about their morphology, including 
whether or not they display arc and multiple image geometries characteristic of 
lensing. For this diagnosis, sub-arcsecond mm/sub-mm images using the 
high-resolution extremely-sensitive interferometer array ALMA,\cite{B_ALMA} or 
very deep radio images using the VLA will be required. Accurate positions for {\it
Planck} sources will first be required: these could be obtained using either ALMA or 
the 3.5-m ESA cornerstone {\it FIRST} telescope.\cite{B98} Note that it is very unlikely 
for the foreground lens galaxy to be bright in the sub-mm waveband,\cite{B98} as this 
would require the foreground lens itself to be a very luminous infrared galaxy, 
and the space density of such sources is rather low. Hence, only the lensed 
images will be detected at bright flux levels in an ALMA continuum image, free from 
the blending and masking effects of emission from the foreground lens. Since the 
first discussion of a {\it Planck} lens survey,\cite{B96} the specifications of ALMA 
have been significantly refined.\cite{Wootten} At sub-mm flux densities of order 
100\,mJy, ALMA will be able to detect galaxies in integrations lasting much less than 
a second, and to make high-quality images in only several minutes.  ALMA
will make it easy to confirm lensing features and obtaining good quality images of 
the galaxies detected in a {\it Planck} survey. 

\begin{figure}
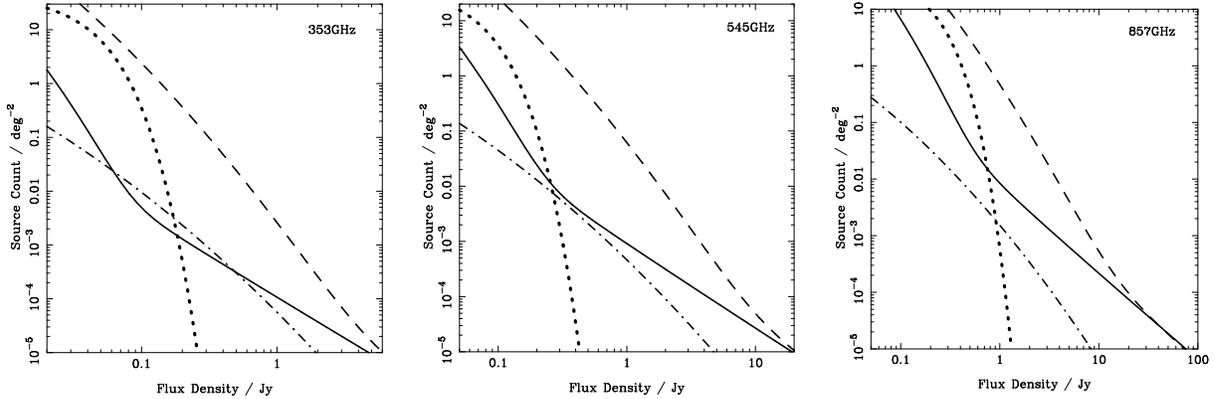

\psfig{figure=NS_Moriond_353.ps,height=1.97in,angle=-90}
\hskip 10pt
\psfig{figure=NS_Moriond_545.ps,height=1.97in,angle=-90}
\hskip 10pt
\psfig{figure=NS_Moriond_857.ps,height=1.97in,angle=-90}
\caption{Examples of counts of galaxies that will be probed in the 
{\it Planck} survey. The solid line shows the count of 
galaxies expected in a model of hierarchical merging galaxies.$^9$ The 
dashed line shows the count of galaxies expected in a model of luminosity 
evolution based on the {\it IRAS} luminosity function.$^8$ 
The dot-dashed line shows the expected count of lensed galaxies, derived in 
the second model.$^7$ The dotted line shows the surface 
density of pixels expected to exceed each flux density due to the effects of source 
confusion, derived by extrapolating the confusion noise distributions shown by the 
solid curves in Fig.\,1. These correspond to the 
second source population, and are thus likely to be a conservative, pessimistic 
estimate of the true confusion noise. To obtain reliable, unconfused detections, the 
counts of sources being sought must lie above the dotted curves in each 
figure.} 
\end{figure}

The form of the mid-infrared SED of the galaxies detected using {\it Planck} can be 
determined using the SPIRE and PACS instruments aboard {\it FIRST}, allowing the 
redshift/dust temperature of the detected galaxies to be determined. Direct 
spectroscopic redshifts for the lensed galaxies should come quite easily from 
molecular emission line data taken using ALMA or from observations of features in 
the redshifted mid-infrared spectrum taken using {\it FIRST}-PACS.\cite{Lutz} 
Redshifts for the lensing galaxies should be readily obtained from the 
frequencies of CO absorption lines imposed on the bright dust continuum emission 
of the magnified background galaxies due to the interstellar medium in 
the foreground lens, a technique which has already been developed and demonstrated 
using existing mm-wave interferometers.\cite{CW}

Other relevant information can be provided by the FIRST 
radio survey at the VLA, which covers over 5000\,deg$^2$ to a depth of 
about 1\,mJy at 1.4\,GHz. The high-resolution radio 
images from FIRST should detect a significant fraction of the {\it Planck} detected 
galaxies with sub-mm 
flux densities of several hundred mJy, based on our current knowledge of the 
SEDs of confirmed high-redshift SCUBA galaxies. 
For example, the 25-mJy 850-$\mu$m sub-mm source 
SMM\,J02399$-$0136 has a 1.4-GHz flux of 520\,$\mu$Jy.\cite{I+7} The 
brighter cousins of these objects in the {\it Planck} survey might typically 
appear at the faintest levels in the FIRST/VLA survey.  The 25-mJy 60-$\mu$m 
all-sky image from the {\it IRIS/ASTRO-F} satellite\cite{Tak} will also be useful 
for detecting {\it Planck} sources.

\section{Summary} 

The {\it Planck Surveyor} CMB imaging mission should detect many thousands 
of high redshift dusty galaxies, a significant fraction of which are expected to 
be gravitationally lensed. In order to make more detailed predictions it will 
be necessary to better quantify the abundance of sub-mm-selected galaxies 
with flux densities of order 100\,mJy. By combining the {\it Planck} results 
with observations made using {\it FIRST} and ALMA a large catalogue of lenses 
should be identified, providing a unique sample of objects for 
further study. The properties of the individual galaxies in the catalogue will 
be useful for obtaining information about both the geometry of the Universe and the  
evolution of distant dusty galaxies.

\section*{Acknowledgements}
The author, Raymond and Beverly Sackler Foundation Research Fellow, 
gratefully acknowledges generous support from the Foundation 
as part of their Deep Sky Initiative Programme at the IoA. Attendance at the meeting 
was made possible by the European Union TMR 
programme, which supports European research in gravitational lensing via  
contract number ERBFMRXCT98-0172 (Lensnet). This work was carried out 
within the Cambridge Planck Analysis Centre (CPAC), supported by PPARC. I thank 
Vicki Barnard, Dave Frayer, Ben Metcalf,  
Daniel Mortlock and members of the {\it FIRST-Planck} extragalactic science team 
for useful discussions and Kate Quirk for comments on 
the manuscript.


\section*{References}
\begin{thebibliography}{99}

\bibitem{Bersanelli}M. Bersanelli {\it et al}, SCI(96)3, (ESA, 
Paris, 1996). 

\bibitem{B96} A.W. Blain, \Journal{MNRAS}{283}{1340}{1996}.

\bibitem{B98} A.W. Blain, \Journal{MNRAS}{297}{511}{1998}.  

\bibitem{B_Carnegie} A.W. Blain, in {\em Photometric Redshifts}, 
ed. R. Weymann {\it et al}, (ASP, San Francisco, 2000). 

\bibitem{B_ALMA} A.W. Blain, in {\em Science with ALMA}, ed. A. Wootten, 
(ASP, San Francisco, 2000).

\bibitem{BIS} A.W. Blain, R.J. Ivison and I. Smail, 
\Journal{MNRAS}{296}{L29}{1998}. 

\bibitem{BMM} A.W. Blain, O. M\"oller and A.H. Maller, 
\Journal{MNRAS}{303}{423}{1999}.

\bibitem{BSIK} A.W. Blain, I. Smail, R.J. Ivison and J.-P. Kneib, 
\Journal{MNRAS}{302}{632}{1999}. 

\bibitem{BJSLKI} A.W. Blain, A. Jameson, I. Smail, M.S. Longair 
{\it et al}, \Journal{MNRAS}{309}{715}{1999}. 

\bibitem{Carilli2000} C.L. Carilli {\it et al}, conference proceeding 
(astro-ph/9907436).

\bibitem{Boulanger} F.X. D\'esert, F. Boulanger and J.L. Puget, 
\Journal{A\&A}{237}{215}{1990}. 

\bibitem{Dunne} L. Dunne {\it et al}, \Journal{MNRAS}{315}{115}{2000}. 

\bibitem{NewFrayer} D.T. Frayer, I. Smail, R.J. Ivison {\it et al},
\Journal{ApJ}{in press}{astro-ph/0005239}{2000}.

\bibitem{Hobson} M.P. Hobson {\it et al}, \Journal{MNRAS}{306}{232}{1999}. 
 
\bibitem{I+7} R.J. Ivison, I. Smail, J.F. Le Borgne, A.W. Blain
{\it et al}, \Journal{MNRAS}{298}{583}{1998}. 

\bibitem{Cloverleaf} J.-P. Kneib, D. Alloin, Y. Mellier {\it et al},
\Journal{A\&A}{329}{827}{1998}. 

\bibitem{APM} G.F. Lewis, S.C. Chapman, R.A. Ibata {\it et al}, 
\Journal{ApJ}{505}{L1}{1998}. 

\bibitem{Lutz} D. Lutz {\it et al}, \Journal{ApJ}{505}{L103}{1998}. 

\bibitem{Papa} P.P Papadopoulos, H.J.A. R\"ottgering, P.P. van 
der Werf {\it et al}, \Journal{ApJ}{528}{626}{2000}. 

\bibitem{10214} M. Rowan-Robinson {\it et al}, 
\Journal{Nature}{351}{719}{1991}. 

\bibitem{SFD} D.J. Schlegel, D.P. Finkbeiner and M. Davis, \Journal{ApJ}
{500}{525}{1998}. 

\bibitem{Solomon} P.M. Solomon, D. Downes, S.J.E. Radford and J.W. 
Barrett, \Journal{ApJ}{478}{144}{1997}.  

\bibitem{Tak} T.T. Takeuchi {\it et al}, \Journal{PASP}{111}{288}{1999}. 

\bibitem{Toffolatti} L. Toffolatti {\it et al},  
\Journal{MNRAS}{297}{117}{1998}. 

\bibitem{CW} T. Wiklind and F. Combes, \Journal{Nature}{379}{139}{1996}. 

\bibitem{Wootten} A. Wootten ed. {\em Science with ALMA}
(ASP, San Francisco, 2000).

\end{thebibliography}
\end{document}